\documentclass[letterpaper,12pt]{article}
\usepackage{amssymb}

\usepackage{ifpdf}
\ifpdf
	\usepackage[pdftex]{graphicx}
	\usepackage[pdftex,unicode,implicit]{hyperref}

	\hypersetup{
  	pdftitle     = {},
  	pdfkeywords  = {},
  	pdfauthor    = {},
  	pdfcreator   = {pdf\LaTeXe\ with package \flqq hyperref\frqq},
  	pdfproducer  = {pdf\LaTeXe\ with package \flqq hyperref\frqq},
  	pdfpagemode  = UseNone,  %change this to FullScreen if you want...
  	pdffitwindow = true,  % Open the pdf file without the sidebars open.
  	unicode      = true,
  	plainpages   = true,
  	colorlinks   = true,  % links have no different colors..
  	citecolor    = blue,  
  	urlcolor     = blue,
  	linkcolor    = blue
	}

	\newcommand{\grqc}[1]{
		{\href{http://www.arXiv.org/abs/gr-qc/#1}{arXiv:gr-qc/#1}}}

	\newcommand{\arXiv}[1]{
		{\href{http://www.arXiv.org/abs/#1}{arXiv:#1}}}

\else
  \usepackage[dvips]{graphicx}
  \usepackage[unicode,implicit]{hyperref}
  
  \newcommand{\grqc}[1]{arXiv:{gr-qc/#1}}
  
  \newcommand{\arXiv}[1]{{arXiv:#1}}

\fi

\makeatletter
\@addtoreset{equation}{section}
\makeatother

\def\hg{\hat{g}}
\def\hp{\hat{\pi}}
\def\sg{\mathrm{sgn}}

\begin{document}

%%%%%%%%%%%%%%%%%%%%%%%%%%%%%%%%%%%%%%%%%%%%%%%%%%%%%%%%%%%%%%%%%%%%%%
%%%%%%%%%%%%%%%%%%%%%%%%%%%%%%%%%%%%%%%%%%%%%%%%%%%%%%%%%%%%%%%%%%%%%%
%%%%%%%%%%%%%%%%%%%%%%%%%%%%%%%%%%%%%%%%%%%%%%%%%%%%%%%%%%%%%%%%%%%%%%

\begin{flushright}
{\small
May $21^{\rm th}$, $2012$}
%\normalsize
\end{flushright}

\begin{center}

\vspace*{2cm}
{\bf \LARGE 
Non-perturbative analysis of the constraints and the positivity of the energy of the complete Ho\v{r}ava theory} 
\vspace*{3em}

%authors
{\sl\large Jorge Bellor\'{\i}n,}$^{a,}$\footnote{\tt jorgebellorin@usb.ve}
{\sl\large Alvaro Restuccia}$^{a,b,}$\footnote{\tt arestu@usb.ve}
{\sl\large and Adri\'an Sotomayor}$^{c,}$\footnote{\tt asotomayor@uantof.cl}
\vspace{3ex}

$^a${\it Departamento de F\'{\i}sica, Universidad Sim\'on Bol\'{\i}var, Valle de Sartenejas,\\ 
1080-A Caracas, Venezuela.} \\[1ex]
$^b${\it Department of Physics, Universidad de Antofagasta, Chile.}
\\[1ex]
$^c${\it Departamento de Matem\'aticas, Universidad de Antofagasta, Chile.}

\vspace*{3em}
{\bf Abstract}
\begin{quotation}{\small
We perform a non-perturbative analysis to the Hamiltonian constraint of the lowest-order effective action of the complete Ho\v{r}ava theory, which includes a $(\partial_i \ln N)^2$ term in the Lagrangian. We cast this constraint as a partial differential equation for $N$ and show that the solution exists and is unique under a condition of positivity for the metric and its conjugate momentum. We interpret this condition as the analog of the positivity of the spatial scalar curvature in general relativity. From the analysis we extract several general properties of the solution for $N$: an upper bound on its absolute value and its asymptotic behavior. In particular, we find that the asymptotic behavior is different to that of general relativity, which has consequences on the evolution of the initial data and the calculus of variations. Similarly, we proof the existence and uniqueness of the solution of the equation for the Lagrange multiplier of the theory. We also find a relationship between the expression of the energy and the solution of the Hamiltonian constraint. Using it we prove the positivity of the energy of the effective action under consideration. Minkowski spacetime is obtained from Ho\v{r}ava theory at minimal energy.
}\end{quotation}

\end{center}

\thispagestyle{empty}

%%%%%%%%%%%%%%%%%%%%%%%%%%%%%%%%%%%%%%%%%%%%%%%%%%%%%%%%%%%%%%%%%%%%%%%%%
%%%%%%%%%%%%%%%%%%%%%%%%%%%%%%%%%%%%%%%%%%%%%%%%%%%%%%%%%%%%%%%%%%%%%%%%%
%%%%%%%%%%%%%%%%%%%%%%%%%%%%%%%%%%%%%%%%%%%%%%%%%%%%%%%%%%%%%%%%%%%%%%%%%
%%%%%%%%%%%%%%%%%%%%%%%%%%%%%%%%%%%%%%%%%%%%%%%%%%%%%%%%%%%%%%%%%%%%%%%%%
\newpage
\section{Introduction}
There are many interesting analyses that can be done in the framework of the Hamiltonian formulation of the Ho\v{r}ava theory \cite{Horava:2009uw}. Such studies are facilitated by the fact that the theory is originally formulated in Arnowitt-Deser-Misner (ADM) variables. In this context, the closeness of Dirac's procedure for the preservation in time of the constraints is of central importance for the consistency of the theory. This requires to analyze carefully the existence and uniqueness of the solutions of the constraints that emerge from Dirac's procedure in the sense of formal problems of partial differential equations (PDEs). This is our main goal in this paper. In particular, we shall study the existence and uniqueness of the solution of the Hamiltonian constraint of the lowest-order effective model of the nonprojectable Ho\v{r}ava theory. The strategy is to regard this constraint as a PDE for the lapse function. We shall also study general properties of the solutions of the constraints and compare them with known results in general relativity. We are interested in analyzing the asymptotic behavior of the field variables since, because of the reduced gauge symmetry group, the formulation is different (and simpler) to the asymptotic behavior in the canonical formulation of general relativity.

In the ADM formulation of general relativity, the perturbative Hamiltonian constraint is solved for the scalar mode of the transverse sector of the spatial metric \cite{Arnowitt:1962hi}. The reduced Hamiltonian is then formulated for the independent modes, the transverse-traceless tensorial modes. In turn, the ADM energy is determined precisely by the asymptotic behavior of the transverse scalar mode. Therefore, there is a direct relationship between the Hamiltonian constraint and the energy, the solution of the former is the one that contribute to the energy. Since in our analysis we deal with the general solution of the Hamiltonian constraint of the Ho\v{r}ava theory, in this paper we also search for a relationship between the variable fixed by the Hamiltonian constraint and the energy of the Ho\v{r}ava theory. We shall perform the study on nonperturbative grounds. We remark that high-order terms in the potential do not contribute to the energy of the theory since the energy is given by a surface integral at spatial infinity and any high-order term decays faster than $1/r^{3}$. Therefore, the energy of the lowest-order effective theory coincides with the energy of the full theory.

The Ho\v{r}ava theory can be formulated in two disconnected versions: the projectable formulation, in which the lapse function depends only in time and the nonprojectable formulation were the lapse function depends both in space and time. Although both possibilities are compatible with the gauge symmetry of the theory, the foliation-preserving diffeomorphisms, the former cannot be recovered from the latter due to the constraints of the nonprojectable case \cite{Bellorin:2010je}. In the projectable version there is not any version of local Hamiltonian constraint (which, in general relativity, is part of the Einstein equations), hence general relativity cannot be recovered in the projectable formulation unless radical modifications to the theory are imposed, as changing, for example, the gauge symmetry group (see \cite{Horava:2010zj}). Because of this obstruction we incline ourselves to the nonprojectable version, where there is a clear scenario in which general relativity emerges \cite{Bellorin:2010je}.

The Hamiltonian formulation of the lowest-order (second-order) truncation of the original Ho\v{r}ava theory was achieved by two of us in Ref.~\cite{Bellorin:2010je}. With \emph{original} we mean that the potential is written only in terms of the curvature tensors and cosmological constant\footnote{The model in \cite{Bellorin:2010je} does not include cosmological constant.}, as was formulated originally by P. Ho\v{r}ava in Ref.~\cite{Horava:2009uw}. However, terms that depends explicitly in the spatial derivatives of the lapse function can be included in the potential through the acceleration vector $a_i = \partial_i \ln N$, which is covariant under foliation-preserving diffeomorphisms. These terms were formally proposed for the Ho\v{r}ava theory in Ref.~\cite{Blas:2009qj}. Assuming as a principle that a complete, renormalizable, theory must include all the terms compatible with the gauge symmetry group, we consider the \emph{complete} Ho\v{r}ava models as those that include the $a_i$ terms.

In Ref.~\cite{Bellorin:2010je} it was found the rather surprising result that the lowest-order truncation of the original theory has a dynamics completely equivalent to general relativity, regardless of the value of the constant $\lambda$ that arises in the kinetic term of the model. In a subsequent study \cite{Bellorin:2010te}, it was analyzed a model with a $R^2$ term in the potential, showing the closure of the algebra of constraints with the presence of an extra, odd, physical mode. The odd dimensionality of the extra mode was previously pointed out in Ref.~\cite{Blas:2009yd} (see also \cite{Li:2009bg} for early evidence)\footnote{Some perturbative analyses of the original nonprojectable theory can be found in Refs.~\cite{Blas:2009yd, varios:perturbations} and for the complete theory in \cite{Blas:2009qj, varios:perturbativeblasmodel}. For a perturbative analysis in the projectable case see, for example, Ref.~\cite{Bogdanos:2009uj}.}.

For the complete theory, the Hamiltonian analysis of the lowest-order effective action \cite{Kluson:2010nf,Donnelly:2011df,Bellorin:2011ff} showed a consistent structure of constraints. In particular, the model has the momentum constraint as the first-class constraint and the Hamiltonian, together with the vanishing of the momentum conjugated to the lapse function $N$, as the second-class constraints\footnote{In the complete Ho\v{r}ava theory, unlike general relativity, the lapse function must be considered as a canonical variable.}. In addition, the preservation in time of the second-class constraints imposes a elliptic PDE for a Lagrange multiplier of the theory. Among the set of constraints, the main difference with general relativity is that the Hamiltonian constraint depends on the lapse function (as in the case of the original theory with high-order curvature terms \cite{Bellorin:2010te,Blas:2009yd}). This suggest that this constraint can be solved for $N$ on general grounds, since it is a linear, elliptic, PDE for $N$. We note that in (nonperturbative) general relativity there is not any plan to solve the Hamiltonian constraint for a specific canonical variable that works consistently in general. In this regard the situation looks better in Ho\v{r}ava theory, but at the same time elucidating the existence (and uniqueness) of the solutions of the Hamiltonian constraint becomes an urgent task for the consistency of the theory. 

We remark that it is only after all the constraints have been solved that one can known where are the independent physical degrees of freedom of the theory. This information is crucial in order to study further physical properties of the propagating modes. For example, in the perturbative analyses that have been done in the complete theory the lapse function is excluded from the independent modes while the extra mode is parameterized in terms of the transverse scalar of the metric and its conjugate momentum. This procedure is implicitly supported on the assumption that the lapse function can be consistently solved from the Hamiltonian constraint, which is the only constraint where it gets involved. Any failure in solving the Hamiltonian constraint for the lapse function could be the signal of a different independent mode and could change drastically the conclusions about the physics of it. Therefore, the rather non trivial structure of the PDE for $N$ that yields the Hamiltonian constraint gives rise relevance to the analysis of the existence of its solutions.

As we mentioned at the beginning, in this paper we elaborate on this point, showing rigorously that the solution, in the sense of distributions, exists and is unique. To arrive at this conclusion we assume certain condition of positiveness for the gravitational variable and its conjugate momentum. This assumption can be seen as the extension of the assumption of positiveness of the scalar curvature in general relativity, as we are going to see.  We shall also study the existence of the solution of the equation for the Lagrange multiplier of the theory, in this case the exact solution since the operator is simpler. The analysis will also allows us to extract several general properties of the solutions for the lapse function and the Lagrange multiplier. We give in advance that the evolution equations preserve the leading mode (non-fixed by the boundary conditions) of the asymptotic expansion of the initial data; that is, the time derivative of the spatial metric has a fall off faster than the metric itself.

Finally, we shall perform a change of variables by doing a canonical transformation that will allow us to reformulate the solution of the Hamiltonian constraint in terms of certain combination of fields. This combination depends on $N$, hence the solution can still be used to fix $N$. The novelty is that the combination is the one that precisely yields the energy, much like (perturbative) general relativity.

The Hamiltonian we will analyze is also related to the famous Perelman energy function \cite{Perelman:2006un}, although the evolution equations are different. It is then relevant its analysis also from this different point of view.

%%%%%%%%%%%%%%%%%%%%%%%%%%%%%%%%%%%%%%%%%%%%%%%%%%%%%%%%%%%%%%%%%%%%%%%%%
%%%%%%%%%%%%%%%%%%%%%%%%%%%%%%%%%%%%%%%%%%%%%%%%%%%%%%%%%%%%%%%%%%%%%%%%%
%%%%%%%%%%%%%%%%%%%%%%%%%%%%%%%%%%%%%%%%%%%%%%%%%%%%%%%%%%%%%%%%%%%%%%%%%
%%%%%%%%%%%%%%%%%%%%%%%%%%%%%%%%%%%%%%%%%%%%%%%%%%%%%%%%%%%%%%%%%%%%%%%%%
\section{Asymptotics in general relativity}
In order to study the similarities and differences between general relativity and the Ho\v{r}ava theory in the asymptotic behavior of the field variables, we start with a quick summary on the subject in general relativity. We cast the condition of asymptotic flatness by demanding that the coordinate system becomes Cartesian at spatial infinity. This approach is convenient for us because of the connection with the problem of solving the PDEs that arise in the canonical formulation of the Ho\v{r}ava theory.

In the standard (ADM) canonical formulation of general relativity one starts by splitting the space-time into spacelike submanifolds and the time direction. To this end one needs a concrete notion of time, hence a global time function $t$ is introduced. On the basis of this function the space-time is regarded as a foliated manifold $\mathcal{M} = \Sigma \times \mathbb{R}$, where $\Sigma$ represents the three-dimensional spacelike slices of constant $t$ and one also assumes that it is endowed with a Riemannian metric $g_{ij}$. Since in this work we are interested in gravitation as a local theory (specially for isolated systems), we consider $\Sigma$ to be a complete, connected, asymptotically flat Riemannian manifold.

Let $n^\mu$ be the future-directed, unit vector field orthogonal to $\Sigma$ with respect to the space-time metric. Being $n^\mu$ orthogonal to $\Sigma$, its action over $t$ does not yield zero. In addition, the future-directed condition restrict $n^\mu \partial_\mu t$ to be positive. The lapse function $N$ is determined by the foliation and the space-time metric by the formula
\begin{equation}
 N   =  \frac{1}{n^\mu \partial_\mu t} \,.
\label{N}
\end{equation}
Let $t^\mu$ be a vector field satisfying 
\begin{equation}
 t^\mu \partial_\mu t = 1 \,.
\label{tcondition}
\end{equation}
One may decompose $t^\mu$ into its normal and tangent components with respect to $\Sigma$. By using (\ref{N}) in condition (\ref{tcondition}), one obtains that $t^\mu$ takes the form
\begin{equation}
 t^\mu  =  N n^\mu + N^\mu \,,
\label{Ni}
\end{equation}
where $N^\mu$, the shift function, is the part of $t^\mu$ lying over $\Sigma$. If one chooses $t$ as the time coordinate and also takes $t^\mu$ such that the time coordinate is adapted to it, $ t^\mu \partial_\mu = \partial_t$, then the space-time metric takes the well-known ADM form,  \begin{equation}
 ds^2 =
 ( - N^2 + N_i N^i ) dt^2 + 2 N_i dx^i dt + g_{ij} dx^i dx^j \,,
\label{ds}
\end{equation}
where $N_i = g_{ij} N^i$.

It is important to define two more objects associated to the foliation. One is the extrinsic curvature $K_{\mu\nu}$, which basically measures the deviation of $n^\mu$ from itself under parallel transport along $\Sigma$. Since $P_\mu{}^\alpha \equiv \delta_\mu{}^\alpha + n_\mu n^\alpha$ is the projector to $\Sigma$, $K_{\mu\nu} \equiv P_\mu{}^\alpha P_\nu{}^\beta \nabla_\alpha n_\beta$. The spatial components of $K_{\mu\nu}$ constitute a tensor over $\Sigma$ and are given by
\begin{equation}
 K_{ij} = \frac{1}{2N} ( \dot{g}_{ij} - \nabla_{(i} N_{j)} ) \,.
\end{equation}
They essentially play the role of conjugate momentum of $g_{ij}$ in the canonical formulation,
\begin{equation}
 \pi^{ij} = \sqrt{g} ( K^{ij} - g^{ij} K ) \,.
\end{equation}
The other object is the acceleration vector,
\begin{equation}
 a_i = \partial_i \ln N \,,
 \label{ai}
\end{equation}
which is the gradient of $N$ along $\Sigma$. This vector plays a key role in the nonprojectable formulation of the Ho\v{r}ava theory since it enters in the potential \cite{Blas:2009qj}.

The asymptotic behavior of $g_{ij}$ and $K_{ij}$ can be conveniently established in terms of the coordinate system. For $r\rightarrow \infty$, where $r \equiv \sqrt{x^i x^i}$ and $x^i$ become Cartesian at spatial infinity, an asymptotically flat metric tends to the Euclidean metric in the form
\begin{equation}
 g_{ij} = \delta_{ij} + \mathcal{O}(r^{-1}) \,,
\label{gdecay}
\end{equation}
whereas the extrinsic-curvature tensor falls off at the speed
\begin{equation}
 K_{ij} = \mathcal{O}(r^{-2}) \,.
\label{decayk}
\end{equation}
Actually, these are consequences of the formal definition of asymptotic flatness, which roughly consists of the requisite that the metric can be completed at infinity in the same way that Minkowski space-time can be completed by a conformal infinity (see, e.~g.,~\cite{Wald:1984rg} and see \cite{Finster} for a definition of asymptotic flatness for $\Sigma$ that involves condition (\ref{gdecay}) directly).

Condition (\ref{gdecay}) has important consequences in the way energy is defined in general relativity. In this theory there is not a universal expression for a local energy density, such that the energy could be formulated as a volume integral. Instead, the energy, as well as the momentum, are formulated as surface integrals at spatial infinity. Specifically, for asymptotically flat space-time, the ADM energy and momentum of the gravitational field are respectively defined by
\begin{eqnarray}
 E_{\mbox{\tiny ADM}} & = & \oint d\Sigma_i ( \partial_j g_{ij} - \partial_i g_{jj} ) \,,
\label{admenergy}
\\[1ex]
 P^i_{\mbox{\tiny ADM}} & = & -2 \oint d\Sigma_j \pi^{ij} \,.
\end{eqnarray}
It turns out that $E_{\mbox{\tiny ADM}}$ has physically acceptable values, that is, finite and different from zero in general, if $g_{ij}$ behaves at spatial infinity as indicated in (\ref{gdecay})\footnote{For this statement we assume that the spatial derivatives of any mode of order $r^{-\beta}$, $\beta>0$, are of order $r^{-(\beta + 1)}$.}. Similarly, the decay (\ref{decayk}) yields
\begin{equation}
 \pi^{ij} = \mathcal{O}(r^{-2}) \,,
\label{pidecay}
\end{equation}
such that the configurations have finite and in general different from zero momentum.

It is worth stressing that the ADM energy (\ref{admenergy}) of all configurations is determined only by the term of order $r^{-1}$ once $g_{ij}$ is expanded at the asymptotic limit (e.~g., the mass parameter in Schwarzschild space-time). Therefore, when performing calculus of variations, variations of order $\delta g_{ij} = \mathcal{O}(r^{-1})$ allow to compare between points of the phase space of (generically) different energy. On the contrary, any variation of faster decay preserve the value of the energy.

The Hamiltonian of general relativity is given by
\begin{eqnarray}
 H_{\mbox{\tiny GR}}  & = & 
 \int d^3x ( N \mathcal{H}_{\mbox{\tiny GR}} + N_i \mathcal{H}^i_{\mbox{\tiny GR}} ) 
 + E_{\mbox{\tiny ADM}} \,,
\label{grhamiltonian}
\\
 \mathcal{H}_{\mbox{\tiny GR}} & \equiv &
 \frac{1}{\sqrt{g}} ( \pi^{ij} \pi_{ij} - {\textstyle\frac{1}{2}}\pi^2 )
 - \sqrt{g} R \,,
 \label{grhamiltoniancons}
\\
 \mathcal{H}^i_{\mbox{\tiny GR}} & \equiv & 
 - 2 \nabla_j \pi^{ji} \,.
\label{grmomentumcons}
\end{eqnarray}
The ADM energy is included in the Hamiltonian in order to get the correct evolution equations under arbitrary variations, which are considered to be $\delta g_{ij} = \mathcal{O}(r^{-1})$ asymptotically \cite{Regge:1974zd}. For variations with this asymptotic fall off, the variation of the curvature term, $\delta ( - \sqrt{g} R )$, gives rise to a nonvanishing boundary contribution that is exactly $-\delta E_{\mbox{\tiny ADM}}$, hence it cancels completely the variation of the boundary term in (\ref{grhamiltonian}). If the $E_{\mbox{\tiny ADM}}$ term was not included in the Hamiltonian, then the boundary contribution of $\delta ( - \sqrt{g} R )$ would be an obstruction to get any evolution equations. This argument is an elegant way to justify the ADM energy of the gravitational field \cite{Regge:1974zd}. Since the bulk part of the Hamiltonian is a sum of constraints, $\mathcal{H}_{\mbox{\tiny GR}} = \mathcal{H}^i_{\mbox{\tiny GR}} = 0$, when the Hamiltonian is evaluated on any physical configuration it yields the corresponding value of $E_{\mbox{\tiny ADM}}$. Moreover, the $E_{\mbox{\tiny ADM}}$ term of the Hamiltonian can be obtained from the very covariant action by requiring its differentiability under arbitrary variations, $\delta g_{\mu\nu} = \mathcal{O}(r^{-1})$ asymptotically \cite{Hawking:1995fd}. For variations with faster decay there are not boundary contributions coming from the bulk part of the Hamiltonian and $\delta E_{\mbox{\tiny ADM}}$ is automatically zero.

The canonical formalism is particularly suitable to formulate initial-data problems in general relativity. As can be seen from (\ref{grhamiltonian}), in general relativity $N$ and $N_i$  can be dropped out from the phase space since they are the Lagrange multipliers of the constraints $\mathcal{H}$ and $\mathcal{H}^i$. These are first-class constraints, hence $N$ and $N_i$ can be determined by gauge fixing the symmetries of general relativity. Therefore, the initial data consists of an asymptotically flat set $(\Sigma,\: g_{ij}\:, \pi^{ij})$ given at the initial time and subject to the constraints (\ref{grhamiltoniancons}) and (\ref{grmomentumcons}). The evolution equations
\begin{eqnarray}
 \dot{g}_{ij} & = & 
 \frac{2N}{\sqrt{g}} (\pi_{ij} - {\textstyle\frac{1}{2}} g_{ij} \pi ) 
 + 2 \nabla_{(i} N_{j)} \,.
\label{dotggr}
\\
\dot{\pi}^{ij} & = &
- N \sqrt{g} \left( R^{ij} - {\textstyle\frac{1}{2}} g^{ij} R \right)
- \frac{2 N}{\sqrt{g}} \left[ 
     \pi^i{}_k \pi^{kj}
     - {\textstyle\frac{1}{2} \pi \pi^{ij}
     - \frac{1}{4} g^{ij} ( \pi^{kl} \pi_{kl} 
              - \frac{1}{2} \pi^2 ) } \right]
\nonumber \\ & &
+ \sqrt{g} ( \nabla^i \partial^j N - g^{ij} \nabla^2 N )
- 2 \nabla_k N^{(i} \pi^{j)k}
+ \nabla_k ( N^k \pi^{ij}) \,,
\label{dotpigr}
\end{eqnarray}
determine the evolution in time of the initial data once the lapse and shift functions have been chosen; that is, the evolution is given in a particular gauge. The evolution equations automatically preserve the constraints (\ref{grhamiltoniancons}) and (\ref{grmomentumcons}) since these are first-class constraints. The initial data $(\Sigma, g_{ij}, \pi^{ij})$ contains twelve functions over $\Sigma$; the four constraints and the four components of an arbitrary diffeomorphism reduce the number of independent initial functions to four.

The standard assumption \cite{Regge:1974zd} for the asymptotic behavior of the lapse and shift functions is that, in the coordinate system that becomes Cartesian at infinity, they behave as
\begin{eqnarray}
 N & = & 1 + \mathcal{O}(r^{-1}) \,,
\label{asymptoticn}
\\[1ex]
 N_i & = & \mathcal{O}(r^{-1}) \,.
\label{asymptoticni}
\end{eqnarray}
These conditions are read from the Schwarzschild solution and, together with (\ref{gdecay}) and (\ref{pidecay}), allow for the differentiability of the Hamiltonian. However, the structures (\ref{asymptoticn}-\ref{asymptoticni}) are not the most general ones since the phase space must include those configurations obtained by the action of the symmetry group of the theory. Coordinate transformations that do not preserve the foliation do not preserve (\ref{asymptoticn}-\ref{asymptoticni}). In particular, there is the possibility of performing a coordinate transformation that involves the action of the Poincar\'e group at spatial infinity. For the case of a pure, infinitesimal, boost at spatial infinity the lapse function takes the form
\begin{equation}
 N = \beta^i x^i + \mathcal{O}(r^{-1}) \,,
\end{equation}
such that it diverges linearly. This is important in order to establish the compatibility of the asymptotic behavior of the field variables with the evolution equations. From (\ref{dotggr}) and (\ref{dotpigr}) and setting $N_i = 0$ one has, asymptotically,
\begin{equation}
 \dot{g}_{ij} = \mathcal{O}(r^{-1}) \,,
\hspace{2em}
 \dot{\pi}^{ij} = \mathcal{O}(r^{-2}) \,.
\end{equation}
Therefore, the decays (\ref{gdecay}) and (\ref{pidecay}) are preserved.

%%%%%%%%%%%%%%%%%%%%%%%%%%%%%%%%%%%%%%%%%%%%%%%%%%%%%%%%%%%%%%%%%%%%%%
%%%%%%%%%%%%%%%%%%%%%%%%%%%%%%%%%%%%%%%%%%%%%%%%%%%%%%%%%%%%%%%%%%%%%%
%%%%%%%%%%%%%%%%%%%%%%%%%%%%%%%%%%%%%%%%%%%%%%%%%%%%%%%%%%%%%%%%%%%%%%
%%%%%%%%%%%%%%%%%%%%%%%%%%%%%%%%%%%%%%%%%%%%%%%%%%%%%%%%%%%%%%%%%%%%%%
%%%%%%%%%%%%%%%%%%%%%%%%%%%%%%%%%%%%%%%%%%%%%%%%%%%%%%%%%%%%%%%%%%%%%%

\section{Hamiltonian and constraints of the Ho\v{r}ava theory}
In this section we give a summary of the Hamiltonian formulation for the lowest-order effective action of the complete nonprojectable Ho\v{r}ava theory \cite{Kluson:2010nf,Donnelly:2011df,Bellorin:2011ff}. The gauge symmetries of the Ho\v{r}ava theory, which are the foliation-preserving diffeomorphisms, correspond to coordinate transformations of the kind
\begin{equation}
 t' = t'(t) \,,
\hspace{2em}
 \vec{x}\,' = \vec{x}\,'(t,\vec{x}) \,,
\label{coordinatetransformation}
\end{equation}
This class of transformations is a subset of the ones allowed in general relativity. The theory is written in terms of the Arnowitt-Deser-Misner variables, with the novelty that the Lagrangian of the complete theory depends on the spatial vector $a_i$ defined in (\ref{ai}) \cite{Blas:2009qj}. 

The second-order action (without cosmological constant) of the complete theory is given by 
\begin{equation}
 S = 
 \int dt d^3 x \sqrt{g} N ( G^{ijkl} K_{ij} K_{kl} + R 
 + \alpha a_i a^i ) \,,
\end{equation}
where $\alpha$ is an arbitrary coupling constant and
\begin{eqnarray}
K_{ij} & = & \frac{1}{2N} ( \dot{g}_{ij} - 2 \nabla_{(i} N_{j)} ) \,,
\\[1ex]
G^{ijkl} & = &
{\textstyle\frac{1}{2}} \left( g^{ik} g^{jl} + g^{il} g^{jk} \right) 
- \lambda g^{ij} g^{kl} \,.
\end{eqnarray}
Whenever $\lambda \neq {1}/{3}$, the inverse of $G^{ijkl}$ is given by
\begin{equation}
\mathcal{G}_{ijkl} = 
\frac{1}{2} (g_{ik} g_{jl} + g_{il} g_{jk} )
+ \frac{\lambda}{1 - 3\lambda} g_{ij} g_{kl} \,.
\label{inverseg}
\end{equation}

The corresponding Hamiltonian takes the form of a sum of constraints plus boundary terms \cite{Kluson:2010nf, Donnelly:2011df,Bellorin:2011ff},
\begin{eqnarray}
 H  & = & 
 \int d^3x ( N \mathcal{H} + N_i \mathcal{H}^i + \sigma \phi ) 
 + E_{\mbox{\tiny ADM}}
 - 2\alpha \Phi_N  \,,
\label{hamiltonian}
\\
 \mathcal{H} & \equiv &
 \mathcal{G}_{ijkl} \frac{\pi^{ij} \pi^{kl}}{\sqrt{g}} 
 + \sqrt{g} ( - R  +  \alpha ( 2 \nabla_i a^i  + a_i a^i ) )
\label{H} \,,
\\
\mathcal{H}^i & \equiv & - 2 \nabla_j \pi^{ji} + \phi \partial^i N \,,
\label{Hi}
\\
\Phi_N & \equiv & \oint d\Sigma_i \partial_i N \,,
\end{eqnarray}
where $\phi$ is the momentum conjugate to $N$ and $\sigma$ is a Lagrange multiplier. The momentum constraint $\mathcal{H}^i = 0$ is the first-class constraint of the theory whereas $\phi = 0$ and the Hamiltonian constraint $\mathcal{H} = 0$ are the second-class ones. The last term in the momentum constraint (\ref{Hi}) is included in order to this constraint becomes the \emph{full} generator of spatial diffeomorphisms, since the combination $\phi \partial_i N$ serves as generator of these diffeomorphisms on $N$ and $\phi$. Notice that the dependence of the momentum constraint on $N$ vanishes on the constrained phase space, hence the Hamiltonian constraint is the only one where $N$ gets involved once all constraints are imposed. As in general relativity, the ADM energy is included as the boundary term needed to obtain the equations of motion under variations $\delta g_{ij}$ of order $\mathcal{O}(r^{-1})$ asymptotically. Similarly, the flux of $N$ at spatial infinity, $\Phi_N$, cancels a nonzero contribution coming from $\delta ( 2 \alpha N  \nabla_i a^i )$ for variations of $N$ with asymptotic decay $\delta N = \mathcal{O}(r^{-1})$.

To preserve in time the second-class constraints a second-order, linear, PDE for $\sigma$ must be imposed. This equation was found in Ref.~\cite{Bellorin:2011ff} in the form
\begin{eqnarray}
&& \nabla^2 (\sigma / N) + a^i \partial_i (\sigma / N) =
 \mathcal{J} \,,
\label{sigmaeq}
\\[1ex]	
&& \mathcal{J} \equiv
 - \gamma \frac{N^{-1}}{\sqrt{g}} \nabla_i (N^2 \nabla^i \pi)
 +  \frac{N}{\sqrt{g}} \mathcal{G}_{ijkl} [ \nabla^i ( a^j \pi^{kl} )
    + 2 a^i a^j \pi^{kl} ]   \,.
\end{eqnarray}
where $\gamma \equiv (1 - \lambda)/(1- 3\lambda) \alpha$. We can be brought this equation to the form
\begin{equation}
 \partial_i (N\sqrt{g} g^{ij} \partial_j (\sigma / N)) =
 \partial_i \left(
 - \gamma N^2 \nabla^i \pi
 +  N^2 \mathcal{G}^i{}_{klm} a^k \pi^{lm} \right) \,,
\label{sigmaeq2}
\end{equation}
such that the source term is also an exact divergence.

In Ho\v{r}ava theory the initial data is less restricted than in general relativity, but the propagation of the constraints requires the knowledge of the master solution of Eq.~(\ref{sigmaeq}) valid at each instant of time (see Ref.~\cite{Donnelly:2011df}). The shift functions $N_i$ are still the Lagrange multipliers of the momentum constraint $\mathcal{H}^i$, hence they are again fixed by a procedure of gauge-symmetry fixing. The general plan, for which we are going to study its consistency in detail in the next section, is to cast the Hamiltonian constraint $\mathcal{H} = 0$ as a PDE for the lapse function $N$. Therefore, the initial data consists of the set $(\Sigma, g_{ij}, \pi^{ij})$ \emph{subject only to the momentum constraint} $\mathcal{H}^i = 0$. At the initial time, the lapse function $N$ must be solved from the equation $\mathcal{H} = 0$. Then, the evolution of the initial data is dictated by the canonical equations of motion of the theory,
\begin{eqnarray}
\dot{g}_{ij} & = &
  2 N \mathcal{G}_{ijkl} \frac{\pi^{kl}}{\sqrt{g}}
+ 2 \nabla_{(i} N_{j)} \,,
\label{dotg}
\\
\dot{\pi}^{ij} & = &
- N \sqrt{g} \left( R^{ij} - {\textstyle\frac{1}{2}} g^{ij} R \right)
- \frac{2 N}{\sqrt{g}} \left( \pi^i{}_k \pi^{kj}
  - \frac{\lambda}{3\lambda -1} \pi \pi^{ij}
  - \frac{1}{4} g^{ij} \mathcal{G}_{klmn} \pi^{kl} \pi^{mn} \right)
\nonumber \\[1ex] & &
+ \sqrt{g} N \left[
     \nabla^i a^j 
   + (1-\alpha) a^i a^j 
   - g^{ij} \left( \nabla_k a^k   
   + (1 - {\textstyle\frac{1}{2}}\alpha) a_k a^k \right) \right]
\nonumber \\[1ex] & &
- 2 \nabla_k N^{(i} \pi^{j)k}
+ \nabla_k ( N^k \pi^{ij}) \,,
\label{dotpi}
\\
\dot{N} & = & \sigma + N^i \partial_i N \,,
\label{dotn}
\end{eqnarray}
once $N_i$ has been gauge fixed\footnote{$\dot{\phi}$ is automatically zero because of the constraints of the theory.}. As we have mentioned, for the whole propagation it is assumed that Eq.~(\ref{sigmaeq}) is systematically solved for $\sigma$ at each instant of time, starting with the initial time where the initial data is given and $N(0)$ is solved form the Hamiltonian constraint $\mathcal{H}(0) = 0$. This requisite is a consequence of the second-class constraints of the theory, unlike general relativity. In the next section we also are going to study the existence and uniqueness of the solution of Eq.~(\ref{sigmaeq}). Now the initial functions over $\Sigma$ are restricted by the three components of the momentum constraint and there are three independent components of a spatial diffeomorphism. This leaves us with six independent initial functions over $\Sigma$.

%%%%%%%%%%%%%%%%%%%%%%%%%%%%%%%%%%%%%%%%%%%%%%%%%%%%%%%%%%%%%%%%%%%%%%%%%%
%%%%%%%%%%%%%%%%%%%%%%%%%%%%%%%%%%%%%%%%%%%%%%%%%%%%%%%%%%%%%%%%%%%%%%%%%%
%%%%%%%%%%%%%%%%%%%%%%%%%%%%%%%%%%%%%%%%%%%%%%%%%%%%%%%%%%%%%%%%%%%%%%%%%%
%%%%%%%%%%%%%%%%%%%%%%%%%%%%%%%%%%%%%%%%%%%%%%%%%%%%%%%%%%%%%%%%%%%%%%%%%%

\section{Existence and uniqueness of the solution of the Hamiltonian constraint}
\label{sec:existence}
We will now discuss the existence and uniqueness of the solution of the Hamiltonian constraint (\ref{H}) and afterwards of the solution of Eq.~(\ref{sigmaeq2}). We shall consider global, nonperturbative, arguments which will be also useful in the constrained problem in the full Ho\v{r}ava theory. First, it is convenient to rewrite the constraint (\ref{H}) in the following way. Let $N$ be continuous with right and left derivatives, then
\begin{equation}
 N = \sg(N) |N| \,.
\end{equation}
It yields
\begin{equation}
 \partial_i N = \sg(N) \partial_i |N|
\end{equation}
because $\partial_i \sg(N)$ is a Dirac delta distribution on the direction orthogonal to the submanifold $N=0$, but is is multiplied by $|N|$ which is zero on the support of the delta distribution. We then have
\begin{equation}
 a_i = \frac{\partial_i N}{N} = \frac{\partial_i |N|}{|N|} 
     = \partial_i \ln |N| \,. 
\end{equation}
The equation $\mathcal{H} = 0$, where $\mathcal{H}$ is given in (\ref{H}), may be formulated as an equation for $|N|$,
\begin{equation}
 - \nabla^2 \sqrt{|N|}
  + \frac{1}{4 \alpha} \left(  R 
  - \mathcal{G}_{ijkl} \frac{\pi^{ij} \pi^{kl}}{g} \right) 
 \sqrt{|N|} = 0 \,.
\label{sqrtbarn}
\end{equation}
Hence, besides proving the existence and uniqueness of the solution one has to show positiveness of it. It will be convenient to write $\sqrt{|N|} = 1 + u$, then Eq.~(\ref{sqrtbarn}) becomes
\begin{equation}
  - \partial_i (\sqrt{g} g^{ij} \partial_j u )  + G u =
  - G \,,
\hspace{2em}
  G \equiv 
  \frac{1}{4\alpha} \left( \sqrt{g} R 
  - \mathcal{G}_{ijkl} \frac{\pi^{ij} \pi^{kl}}{\sqrt{g}} \right) \,.
\label{ueq}
\end{equation}

Let $u$ be a scalar function on $\Sigma$. We introduce the space $\hat{C}^{1}(\Sigma)$ according to
\begin{equation}
 \hat{C}^{1}(\Sigma) = 
 \{ u \in C^{1}(\Sigma) \: : \: u \rightarrow 0 \:\: \mbox{when} 
                             \:\: r \rightarrow \infty \} \,.
\end{equation}
$C^{1}(\Sigma)$ is the space of functions that have continuous derivatives. It is invariant under diffeomorphisms on the slice $\Sigma$. 

We denote, for $u$ and $v$ in $\hat{C}^{1}(\Sigma)$,
\begin{equation}
 (u,v) = 
 \int\limits_\Sigma d^3 x ( \sqrt{g} g^{ij} \partial_i u \partial_j v + G u v ) \,, 
\label{escalarproduct}
\end{equation}
where $G$ is defined in (\ref{ueq}). The map (\ref{escalarproduct}) is a bilinear functional on $\Sigma$. In the definition of this map we assume that $R$ is integrable over $\Sigma$. If $ G \geq 0 $ on $\Sigma$, then $(u,u) \geq 0$ and $(u,u) = 0$ if and only if $u=0$. In fact, $(u,u) = 0$ implies $\int d^3x \sqrt{g} g^{ij} \partial_i u \partial_j u = 0$, hence $u = \mbox{constant}$ on $\Sigma$ and the boundary condition implies that $u=0$ on $\Sigma$. Consequently, if $G\geq 0$ on $\Sigma$, $(u,v)$ defines an internal product on $\hat{C}^{1}(\Sigma)$.

Let $\hat{\mathcal{H}}^{1}$ denote the Hilbert space obtained by the completion of $\hat{C}^{1}(\Sigma)$ with respect to the norm induced by the internal product (\ref{escalarproduct}). We can now prove the following\\
\noindent {\bf Proposition 1}: Given a complete, asymptotically flat Riemannian manifold $\Sigma$, momenta $\pi^{ij}$ over it and values of the parameters $\lambda$ and $\alpha$ satisfying $G\geq 0$, there always exists in $\hat{\mathcal{H}}^{1}$ a unique weak solution $u$ of Eq.~(\ref{ueq}).\\
\noindent \underline{Proof}: In order to prove this statement  may use the Lax-Milgram theorem. However, in this case it is a straightforward application of the Riesz theorem. In fact, for any $v \in \hat{\mathcal{H}}^{1}$, $\int_\Sigma G v$ is a linear, continuous, functional with domain $\hat{\mathcal{H}}^{1}$. We notice that $G v$ is integrable at infinity. Riesz's theorem ensures the existence of a unique $w \in \hat{\mathcal{H}}^{1}$ such that 
\begin{equation}
 ( w , v ) = - \int\limits_\Sigma G v 
\label{riesz}
\end{equation}
for all $v \in \hat{\mathcal{H}}^{1}$. But Eq.~(\ref{riesz}) is exactly Eq.~(\ref{ueq}) for $w$ when the latter is applied as a distribution to a test function $v \in \hat{\mathcal{H}}^{1}$ with compact support. In fact, $\sqrt{g} g^{ij} \partial_i w \partial_j v$ and $G v$ are scalar densities on $\Sigma$ that are integrable at infinity for all $v \in \hat{\mathcal{H}}^{1}$, in particular, for $C^{\infty}$ test functions of compact support. We then have a unique solution of Eq.~(\ref{ueq}) in the sense of distributions.

\noindent {\bf Remark}: The assumption $G\geq 0$ is essential to prove existence and uniqueness of the solution. If $G$ is bounded from below, $G \geq -C$, $C > 0$, then in a generic situation $ - \nabla^2 + G$ has a nontrivial kernel $K$. In this case Eq.~(\ref{ueq}) has a solution if and only if \emph{$G$ is orthogonal to the kernel $K$}. The solution, if exists, is the addition of a particular solution plus the zero modes. It is then not unique and it may even not exists. 

We can improve the result, as usual in the theory of elliptic operators, to show that the solution is $C^{\infty}$ if $g_{ij}$ and $\pi^{ij}$ are $C^{\infty}$. To do so, we first characterize the asymptotic behavior of the solution.

\noindent {\bf Proposition 2}: Given an asymptotically flat Riemannian manifold with $C^{\infty}$ metric $g_{ij}$ and momenta $\pi^{ij}$, then the asymptotic solution of Eq.~(\ref{ueq}) exists, it is $C^\infty$ for large enough $r$ and its asymptotic behavior is $u =\mathcal{O}(r^{-1})$.\\
\noindent \underline{Proof}: The asymptotic version of Eq.~(\ref{ueq}) is equivalent to perturbate it around the Minkowski solution. Since the momentum $\pi^{ij}$ enters quadratically in Eq.~(\ref{ueq}), it dissapears from this equation at linear order. Hence, for our purposes it is enough to deal with the perturbative metric and $u$. Let us expand the metric at the asymptotic limit,
\begin{equation}
 g_{ij} = \delta_{ij} + h_{ij} \,,
\label{perturbation}
\end{equation} 
such that $h_{ij} = \mathcal{O}(r^{-1})$. At linear order in $h_{ij}$, the scalar curvature takes the form
\begin{equation}
 R = \partial_{ij} h_{ij} - \partial_{ii} h \,,
\label{linearr}
\end{equation}
where $h \equiv h_{jj}$. We use the shorthand $\partial_{ij} = \partial_i \partial_j$, and so on. As usual in perturbation theory, we decompose the perturbative metric into transverse and longitudinal parts,
\begin{equation}
 h_{ij} = 
 h_{ij}^{TT} 
 + {\textstyle\frac{1}{2}}[ \delta_{ij} 
    - \partial_{ij} (\partial_{kk})^{-1}] h^T
 + \partial_i h^L_j + \partial_j h^L_i 
 + \partial_{ij} (\partial_{kk})^{-1} h^L \,,
\label{decomposition}
\end{equation}
where $h^{TT}_{ii} = 0$, $\partial_i h^{TT}_{ij} = 0$, $\partial_i h^L_i = 0$ and the boundary conditions are used in the definition of $(\partial_{kk})^{-1}$. All of these variables are determined uniquely in terms of $h_{ij}$. The scalar of the longitudinal sector is given by
\begin{equation}
 h^L = (\partial_{kk})^{-1} \partial_{ij} h_{ij} \,,
\end{equation}
and starting from this the other variables are solved: the longitudinal vector $h^L_i$ is solved from the divergence of Eq.~(\ref{decomposition}), $\partial_j h_{ij} = \partial_{jj} h^L_i + \partial_i h^L$, the transverse scalar is obtained from $h = h^T + h^L$, and $h^{TT}_{ij}$ can be solved directly from (\ref{decomposition}). After decomposition (\ref{decomposition}), the linearized scalar curvature (\ref{linearr}) is determined only by scalar of the transverse sector,
\begin{equation}
 R = - \partial_{ii} h^T \,.
\end{equation}
We now expand the Eq.~(\ref{ueq}) at $r\rightarrow\infty$. We notice that among the terms that depend on $u$ the dominant one is the Laplacian. In the right hand side, the dominant term in $G$ is $R$. Thus, for large enough $r$, the Eq.~(\ref{ueq}) takes the linearized form
\begin{equation}
 \partial_{ii} u = \partial_{ii} h^T \,.
\end{equation}
The solutions to this equation exist and are given by $u = h^T$ plus an element of the kernel of $\partial_{ii}$. The boundary condition $u|_\infty = 0$ fixes the solution $u = h^T$. Clearly, the solution is $C^\infty$ provided that the metric is $C^{\infty}$ and near spatial infinity it behaves as $u = h^T = \mathcal{O}(r^{-1})$.

Now we are going to show that, as a consequence of Proposition 2, the weak solution of Proposition 1 is asymptotically $C^{\infty}$.

\noindent {\bf Proposition 3}: Given a complete, asymptotically flat Riemannian manifold $\Sigma$ with $C^{\infty}$ metric and conjugated momenta such that $G \geq 0$ on $\Sigma$, the weak solution of Proposition 1 is $C^{\infty}$ on $\Sigma$ and has the asymptotic behavior of Proposition 2.\\
\noindent \underline{Proof}: For the proof we use the elliptic regularity theorem. Under the assumptions $G \in H^{\mbox{\tiny loc}}_s$  for all $s$, hence the weak solution belongs to $H^{\mbox{\tiny loc}}_{s+2}$ for all $s$. We then conclude that $u \in C^{\infty}$ and it has the asymptotic behavior of Proposition 2.

\noindent {\bf Proposition 4}: Under the same assumptions of Proposition 3 and assuming also that $\Sigma$ is path connected, the solution of Eq.~(\ref{ueq}) satisfies $1 + u \geq 0$.\\
\noindent\underline{Proof}: We proceed by contradiction. Let us suppose that at some point $p$ on $\Sigma$ the solution satisfies $u(p) < -1$. At spatial infinity $u \rightarrow 0$, hence, by continuity, on a path form $p$ to infinity there exists a point at which $u = -1$. We denote $u^{-1}(a)$ the set of points of $\Sigma$ at which the solution is $u = a$, and also $M_a = \{ q\in \Sigma \: : \: 1 + u(q) \leq a \}$. If $-1$ is not a critical value of $u$, that is, $||\partial_i u|| \neq 0$ for all $q \in u^{-1}(-1)$, then the implicit function theorem ensures that $u^{-1}(-1)$ is a submanifold of $\Sigma$. Moreover, $u^{-1}(-1)$ is closed and bounded. Since $\Sigma$ is complete then $M_0$ is compact with boundary $\partial M_0 = u^{-1}(-1)$. From Eq.~(\ref{ueq}) we obtain on $M_0$:
\begin{equation}
 \int\limits_{\partial M_0} d\sigma n_i g^{ij} \sqrt{g} ç
   (1+u) \partial_j (1+u) =
 \int\limits_{M_0} d^3x  \left[
     \sqrt{g} g^{ij} \partial_i ( 1 + u ) \partial_j ( 1 + u )
   + G (1 + u )^2 \right] \,.
\end{equation}
But at $\partial M_0$ we have $1 + u = 0$, thus
\begin{equation}
 \int\limits_{M_0} d^3x  \left[
     \sqrt{g} g^{ij} \partial_i ( 1 + u ) \partial_j ( 1 + u )
   + G (1 + u )^2 \right] = 0 \,,
\end{equation}
which yields $1 + u = \mbox{constant}$. Since at $\partial M_0$ it holds that $1 + u = 0$, then $1 + u = 0$ everywhere on $M_0$. If $-1$ is a critical value of $u$, we consider $u^{-1} (-1 + \epsilon)$ and use the continuity of the integral in $M_\epsilon$ as $\epsilon \rightarrow 0$ to conclude, by the same arguments, that $ 1 + u = 0$ on $M_0$. We thus reach a contradiction with the assumed condition $u(p) < -1$ in each case. Therefore, we have $1 + u \geq 0$ and we can identify $|N| = 1+u$.

\noindent {\bf Proposition 5}: Under the assumptions of proposition 4, the solution of the Eq.~(\ref{ueq}) satisfies $|N| \leq 1$.\\
\noindent\underline{Proof}: Let us suppose that $|N| = a > 1$ at some point $p$ on $\Sigma$. Then the set $\hat{M}_a \equiv \{ p \in \Sigma \: : \: |N| \geq a \}$ is closed and bounded and, since $\Sigma$ is complete, it is compact. Then Eq.~(\ref{sqrtbarn}) implies
\begin{equation}
 \frac{1}{2} \int\limits_{\partial \hat{M}_a} 
    n_i g^ {ij} \sqrt{g} \partial_j |N| =
 \int\limits_{\hat{M}_a} (
   \sqrt{g} g^{ij} \partial_i \sqrt{|N|} \partial_j \sqrt{|N|}
   + G |N| ) \,.  
\end{equation}
That is, the outgoing flux of $\partial_i |N|$ through $\partial \hat{M}_a$ is positive. However, from the definition of $\hat{M}_a$ we must have that if $a < b \: \Rightarrow \: \hat{M}_b \subset \hat{M}_a$, which contradicts the flux relation. Consequently $|N| \leq 1$.

\noindent{\bf Remark}: The solution for $N$ may have $+1$ or $-1$ as asymptote. If $\Sigma$ has several disconnected sectors at infinity we may impose the $+1$ or $-1$ value as the asymptotic condition on each of the sectors. The situation is then very different from what occurs in the large-distance effective action studied in Ref.~\cite{Bellorin:2010je}, the $\lambda R$-model, where the constraint for $N$ is
\begin{equation}
 - \partial_i ( \sqrt{g} g^{ij} \partial_j N )
 + \sqrt{g} R N = 0 \,.
\end{equation}
In this case, the conservation of the Hamiltonian constraint,
\begin{equation}
  \mathcal{G}_{ijkl} \frac{\pi^{ij} \pi^{kl}}{\sqrt{g}} 
  -  \sqrt{g} R = 0 \,,
\label{lambdar}
\end{equation}
implies $\pi = 0$, which, when combined with the Hamiltonian constraint, implies that the scalar curvature is nonnegative,
\begin{equation}
 R = \frac{\pi^{ij} \pi_{ij}}{g} \geq 0 \,.
\label{hamiltonianlambdar}
\end{equation}
All the propositions we have considered here apply also for this case. The assumption $G \geq 0$ is now replaced by $R \geq 0$, which now is a consequence of the constraints of the theory. The solution for $N$ is then positive and satisfies $0 \leq N \leq 1$. Note that in general relativity, when the $\pi = 0$ gauge is not imposed, the condition $R \geq 0$ is not implied by the constraints in general.

\noindent{\bf Remark}: If we take the asymptotic boundary condition $|N| \rightarrow 0$ at spatial infinity then $ |N| = 0$ on the whole $\Sigma$. Hence this boundary condition has to be dismissed. We may take $|N| \rightarrow \mbox{constant} \neq 0$ and everything follows in the same way as for $|N| \rightarrow 1$.

\noindent{\bf Remark}: The net outgoing flux of $\partial_i |N|$ is
\begin{equation}
 \frac{1}{2} \int\limits_{\partial\Sigma} 
    n_i g^ {ij} \sqrt{g} \partial_j |N| =
 \int\limits_{\Sigma} (
   \sqrt{g} g^{ij} \partial_i \sqrt{|N|} \partial_j \sqrt{|N|}
   + G |N| ) \geq C \,.  
\label{flux}
\end{equation}
If we take $N \rightarrow 1$ at infinity then $N = |N|$ and the flux of $\partial_i N$ is always positive. We notice that $\sqrt{|N|}$ is not in $\hat{\mathcal{H}}^{1}$, hence the right hand side of Eq.~(\ref{flux}) is not a norm.

The \emph{minimal flux condition} occurs for $N=|N| = 1$. In this case the constraint (\ref{sqrtbarn}) reduces to the Hamiltonian constraint of the $\lambda R$ model (\ref{lambdar}), that is, $G=0$. Since the evolution equation for $N$ is $\dot{N} = \sigma$, Eq.~(\ref{sigmaeq2}) reduces to the corresponding equation $\pi = 0$ of the $\lambda R$ theory. From the evolution equation for $\pi$ we get $R = 0$. From the Hamiltonian constraint (\ref{hamiltonianlambdar}) we then get $\pi^{ij} = 0$. We then get $R_{ij} = 0$ and the three dimensional slices $\Sigma$ are flat. Since $N=1$ we then regain the Minkowski space-time at the minimum flux condition.

\noindent{\bf Proposition 6}: Under he assumptions of proposition 2, the exact solution of Eq.~(\ref{sigmaeq}) exists and is unique. It is $C^{\infty}$ and its asymptotic behavior is $\mathcal{O}(r^{-2})$.\\
\noindent\underline{Proof}: If $N\neq 0$, Eq.~(\ref{sigmaeq}) expresses an elliptic operator acting on $\sigma/N$ equal to a source term. The operator $\partial_i ( N \sqrt{g} g^{ij} \partial_j \: \cdot )$ is strongly elliptic on the complement of the $N=0$ set, which we denote by $M_0$. $M_0$ is compact, on its complement $\bar{M}_0$ the solution $\sigma/N$ of the Eq.~(\ref{sigmaeq}) with boundary condition $\sigma / N = 0$ on $\partial\Sigma$ exists and is unique. At the asymptotic limit $r \rightarrow \infty$ we expand perturbatively the field variables as in (\ref{perturbation}), also expanding the momenta around zero,
\begin{equation}
 \pi^{ij} = p^{ij} \,.
\end{equation}
At linear order Eq.~(\ref{sigmaeq2}) takes the form
\begin{equation}
 \partial_{ii} \sigma = - \gamma \partial_{ii} p \,,
\end{equation}
where $p = p^{ii}$. With the boundary condition, the unique solution to this equation is $\sigma = -\gamma p$, which is $C^\infty$ and of order $\mathcal{O}(r^{-2})$.

Now we are in position to establish the differences in the asymptotic behavior of the field variables between general relativity and Ho\v{r}ava theory. The main point is that in the nonprojectable version of the Ho\v{r}ava theory the lapse function is not a Lagrange multiplier that could be fixed by appealing to any gauge symmetry. Instead, $N$ is determined by the Hamiltonian constraint (\ref{sqrtbarn}). Therefore, the asymptotic behavior of $N$ is dictated by the rest of canonical variables. We have seen that, near spatial infinity, the Hamiltonian constraint (\ref{sqrtbarn}) implies
\begin{equation}
 N = 1 + \mathcal{O}(r^{-1}) \,.
\label{decayn}
\end{equation}
This decay cannot be changed by any coordinate transformation between the kind allowed in Ho\v{r}ava theory (\ref{coordinatetransformation}). This  is in contrast with general relativity, where the asymptotic behavior (\ref{asymptoticn}) of $N$ is understood to hold in a specific coordinate system that becomes Cartesian at infinity, but it is changed by general Lorentz transformations at infinity, in particular by boosts. In Ho\v{r}ava theory the decay (\ref{decayn}) holds for any asymptotically flat configuration and it is preserved by translations and spatial rotations at infinity, which are the only symmetries of the Minkowski space allowed by the theory.

As a consequence, the evolution equations (\ref{dotg}) and (\ref{dotpi}) dictate that the time derivatives of the metric and momentum decay as
\begin{equation}
 \dot{g}_{ij} = \mathcal{O}(r^{-2}) \,,
\hspace{2em}
 \dot{\pi}^{ij} = \mathcal{O}(r^{-3}) \,.
\label{dotgdecay}
\end{equation}
These are faster than the decays of $g_{ij}$ and $\pi^{ij}$ themselves at spatial infinity; in contrast, again, with the observed in general relativity. This feature also holds for $N$, which can be seen from combining the decay of $\sigma$, $\sigma = \mathcal{O}(r^{-2})$, with Eq.~(\ref{dotn}), yielding
\begin{equation}
 \dot{N} = \mathcal{O}(r^{-2})
\label{dotndecay}
\end{equation}
asymptotically. These results imply that in Ho\v{r}ava theory a part of the initial data is always preserved by the evolution equations: the asymptotic modes of order $r^{-1}$ for $g_{ij}(0)$ and $r^{-2}$ for $\pi^{ij}(0)$. Similarly, the mode $r^{-1}$ of $N(0)$ is preserved.

Finally, the results (\ref{dotgdecay}) and (\ref{dotndecay}) forces us to reconsider the functional variations admissible by the Ho\v{r}ava theory, since these variations must belong to the subspace of the fields themselves. The most general asymptotic behavior for variations that preserve (\ref{dotgdecay}) and (\ref{dotndecay}) are
\begin{equation}
 \begin{array}{rclrclrcl}
  \delta g_{ij} & = & \mathcal{O}(r^{-1}) \,, &
\hspace{2em}
  \delta \pi^{ij} & = & \mathcal{O}(r^{-2}) \,, &
\hspace{2em} 
  \delta N & = & \mathcal{O}(r^{-1}) \,,
\\[1ex]
  \delta \dot{g}_{ij} & = & \mathcal{O}(r^{-2}) \,, &
  \delta \dot{\pi}^{ij} & = & \mathcal{O}(r^{-3}) \,, &
  \delta \dot{N} & = & \mathcal{O}(r^{-2}) \,.
\end{array}
\end{equation}

%%%%%%%%%%%%%%%%%%%%%%%%%%%%%%%%%%%%%%%%%%%%%%%%%%%%%%%%%%%%%%%%%%%%%%%%%%
%%%%%%%%%%%%%%%%%%%%%%%%%%%%%%%%%%%%%%%%%%%%%%%%%%%%%%%%%%%%%%%%%%%%%%%%%%
%%%%%%%%%%%%%%%%%%%%%%%%%%%%%%%%%%%%%%%%%%%%%%%%%%%%%%%%%%%%%%%%%%%%%%%%%%
%%%%%%%%%%%%%%%%%%%%%%%%%%%%%%%%%%%%%%%%%%%%%%%%%%%%%%%%%%%%%%%%%%%%%%%%%%

\section{The energy of the theory and the solution of the Hamiltonian constraint}
We consider the conformal transformation
\begin{eqnarray}
 g_{ij} & = & e^{\varphi} \hat{g}_{ij} \,,
\label{reescalingg}
\\[1ex]
 \pi^{ij} & = & e^{-\varphi} \hat{\pi}^{ij} \,,
\end{eqnarray}
where the Riemannian metric $\hg_{ij}$ is subject to
\begin{equation}
\det{\hat{g}_{ij}} = 1
\label{detg}
\end{equation}
(this can be imposed in any coordinate system). We also make the covariant decomposition of $\hp^{ij}$ into its traceless and trace parts with respect to the metric $\hat{g}_{ij}$,
\begin{equation}
 \hat{\pi}^{ij} = \hat{\pi}_T^{ij} + \frac{1}{3} \hat{g}^{ij} \hat{\pi} \,,
\label{pidecomposition}
\end{equation}
where $\hg^{ij}$ is the inverse of $\hat{g}_{ij}$, $\hp \equiv \hg_{ij} \hp^{ij}$ and $\hp_T^{ij}$ is traceless, $\hg_{ij} \hp_T^{ij} = 0$. These variables will allows us to manage the solution of the Hamiltonian constraint of the theory (\ref{H}) and also to connect the solution with the expression of the energy, as we are going to see. The transformation (\ref{reescalingg}-\ref{pidecomposition}) is canonical in the sense that the kinetic terms preserve their canonical structure. Indeed, by imposing condition (\ref{detg}), it is straightforward to obtain
\begin{equation}
 \int\limits_\Sigma d^3x \:\pi^{ij} \dot{g}_{ij} =
 \int\limits_\Sigma d^3x \:( \hp_T^{ij} \dot{\hg}_{ij} + \hat{\pi} \dot{\varphi} ) \,.
\end{equation}

Let us start with the computation of the energy. It is given by the value of the Hamiltonian (\ref{hamiltonian}) on each configuration. Since the bulk part of the Hamiltonian is a sum of constraints, it does not contribute to the energy of any physical configuration. Thus, the energy is given by the values of the boundary terms,
\begin{equation}
 E = E_{\mbox{\tiny ADM}} - 2\alpha \Phi_N \,.
\label{energy}
\end{equation}
This feature will hold even when high-order terms will be included in the potential of the theory. Moreover, since any high-order term has a decay faster that $1/r^3$ at spatial infinity, they do not contribute to the energy of the theory. Therefore, expression (\ref{energy}) is actually the energy of the \emph{full} theory. This one of the aspects that give relevance to the lowest-order effective theory.
In Ref.~\cite{Garfinkle:2011iw} it was formulated a positive-energy theorem for the complete nonprojectable Ho\v{r}ava theory. One of the hypothesis of the theorem is to impose the condition $\pi = 0$.

After the transformation (\ref{reescalingg}-\ref{pidecomposition}), the ADM energy takes the form
\begin{equation}
E_{\mbox{\tiny ADM}} = 
 \oint d\Sigma_i e^\varphi \left( 
     \partial_j \varphi \hg_{ij} 
   + \partial_j \hg_{ij}
   - \partial_i \varphi \hg_{jj}
   + \partial_i \hg_{jj} \right) \,.
\label{prevadmenergy}
\end{equation}
We may further refine this expression if we assume the asymptotic behavior $\hg_{ij} = \delta_{ij} + \mathcal{O}(1/r)$ and $\varphi = \mathcal{O}(1/r)$. It follows that at spatial infinity $\hg_{jj} = 3$ and also, by combining with condition (\ref{detg}), $\partial_i \hg_{jj} = (\det\hg)^{-1} \partial_i \det \hg = 0$. We may also fix a partial gauge in which the longitudinal part of $\hg_{ij}$ vanishes,
\begin{equation}
 \partial_j \hg_{ij} = 0 \,.
\end{equation}
To preserve in time this partial gauge-fixing condition one must combine it with the evolution equation (\ref{dotg}). The resulting equation can be interpreted as the condition fixing $N_i$. After these considerations the ADM energy (\ref{prevadmenergy}) takes the form
\begin{equation}
E_{\mbox{\tiny ADM}} = 
 -2 \oint d\Sigma_i  
     \partial_i \varphi  \,,
\end{equation}
which only depends on $\varphi$. Finally, we get that the energy of the Ho\v{r}ava theory is given by the surface term at spatial infinity
\begin{equation}
E = 
  -2 \oint d\Sigma_i \partial_i ( \alpha \ln N + \varphi )  \,,
\end{equation}
where for convenience we have written $\partial_i N$ as $\partial_i \ln N$ at spatial infinity. Thus, we obtain that the energy of the Ho\v{r}ava theory is determined by the variables $N$ and $\varphi$ in terms of the combination $\xi \equiv \alpha \ln N + \varphi$.

Now let us study the Hamiltonian constraint (\ref{H}). By performing the change of variables (\ref{reescalingg}-\ref{pidecomposition}) into this constraint we obtain
\begin{equation}
\begin{array}{rcl}
 \mathcal{H} & = &
 e^{-3\varphi/2} \left( \hp_T^{ij} \hp_{T\:ij} 
 + (3 - 9\lambda)^{-1}\: \hp^2 \right)
\\[2ex] & &
 +\: e^{\varphi/2} \left( - \hat{R} 
    + 2 \hat{\nabla}^2 \xi
    + \alpha a_i a^i
    + \alpha a_i \partial^i \varphi 
    + \frac{1}{2} \partial_i \varphi \partial^i \varphi \right) \,.
\end{array}
\end{equation}
In this expression indices have been raised/lowered with $\hg_{ij}$ and $\hat{R}$, $\hat{\nabla}$ refer to $\hg_{ij}$. From the $\hp^2$ term of this constraint we notice that, in order to the mode $(\varphi,\hp)$ has a good physical propagation, the value of $\lambda$ must be restricted by $\lambda < 1/3$, which is one of the admissible ranges found in Ref.~\cite{Blas:2009qj}.

As we have seen, the combination $\xi = \alpha \ln N + \varphi$ is the variable that determines the energy of the Ho\v{r}ava theory. On the other hand, in Ho\v{r}ava theory the Hamiltonian constraint is an equation for $N$. These facts motivate us to solve the Hamiltonian constraint for $N$ in terms of the composed variable $\xi$. This can be achieved since in this constraint we can combine all the terms that depend on $N$ in a single elliptic operator acting on $e^{f(\xi)}$, where $f$ is proportional to $\xi$. Indeed, it is straightforward to check the identity
\begin{equation}
 \partial_i (e^{\beta \varphi} \hat{g}^{ij} \partial_j e^{\xi/2\alpha} )
= 
 \frac{1}{4\alpha} e^{\beta\varphi} e^{\xi/2\alpha} \left[
      2 \hat{\nabla}^2 \xi
    + \alpha a_i a^i
    + \alpha a_i \partial^i \varphi 
    + \left( {\textstyle\frac{1}{2}} + \beta \right) \partial_i \varphi \partial^i \varphi 
  \right] \,,
\end{equation}
where $\beta \equiv (\alpha - 2)/2 \alpha$. With this identity the constraint $\mathcal{H} = 0$ can be formulated as the following equation for $\xi$:
\begin{eqnarray}
&& - \partial_i (e^{\beta \varphi} \hat{g}^{ij} \partial_j e^{\xi/2\alpha} )
 + G e^{\xi/2\alpha} = 0 \,,
\label{Hxi}
\\[1ex]
&&  G \equiv - \frac{e^{\beta\varphi}}{4\alpha} \left[
  e^{-2\varphi} \left( \hp_T^{ij} \hp_{T\:ij} 
 + (3 - 9\lambda)^{-1} \: \hp^2 \right)
 - \hat{R} - \beta \partial_i \varphi \partial^i \varphi \right] \,. 
\label{Gxi}
\end{eqnarray}
Note that $G$ depends on the canonical variables $\hg_{ij}$, $\hp_T^{ij}$, $\varphi$ and $\hp$, but does not depend on $N$ and does not have explicit dependence on $\xi$. If $|\varphi|$ is bounded, the composed metric $e^{\beta \varphi} \hg^{ij}$ arising in the differential operator $- \partial_i (e^{\beta \varphi} \hat{g}^{ij} \partial_j \:\cdot )$ is positive definite. Thus, the operator is coercitive. Under the assumption $G \geq 0$, where $G$ is given in (\ref{Gxi}), all the results of Sec.~4 about the existence of the weak solution hold. Therefore, we may solve Eq.~(\ref{Hxi}) for $N$ if we solve it for the combination $\xi$, which is the only place $N$ where arises.

The Ho\v{r}ava theory may be formulated in terms of $\hg_{ij}$, $\hp^{ij}$, $N$. The relation between the Hamiltonian constraint formulated in terms of $\xi$ and the expression of the energy in terms of the flux of $\partial_i \xi$ yields directly the positivity of the energy of the Ho\v{r}ava theory. In fact, from (\ref{Hxi}) we have
\begin{equation}
 E = -2 \oint d\Sigma_i \partial_i \xi = 
 - 4 \alpha \int\limits_\Sigma d^3x \left[
  \frac{1}{4\alpha^2} e^{\xi/\alpha} e^{\beta\varphi}
   \hg^{ij} \partial_i \xi \partial_j \xi
 + G e^{\xi/\alpha} \right] 
\end{equation}
and the right hand member is positive assuming $\alpha < 0$. We remark that the condition $G \geq 0$ is imposed to ensure the existence of of the solution for $\xi$, and, as a consequence, it leads to the positivity of the energy. Moreover, we have shown in a previous remark that $N = 1$ implies, using the field equations, that $g_{ij} = \delta_{ij}$ and $\pi^{ij} = 0$. Since $N=1$ we end up with the Minkowski spacetime, hence $\varphi=0$ and $\xi=0$. Consequently, $N=1$ is a configuration of minimal energy, $E=0$. It remains open the question if there are other configurations of minimal energy.

The condition $\alpha < 0$ implies the same sign in the curvature term $R$ and the $a_i a^i$ term in the Hamiltonian (\ref{hamiltonian}). These terms, together with the factor $N$, correspond exactly to the Perelman energy function
\begin{equation}
 \int\limits_{\Sigma} ( R + \nabla_i f \nabla_i f ) e^{-f} dV \,,
\end{equation}
if we take $f=-\ln N$. The term $\mathcal{G}_{ijkl} \frac{\pi^{ij} \pi^{kl}}{\sqrt{g}}$ in (\ref{H}) implies a different dynamics from the Ricci flow in Perelman's work.

%%%%%%%%%%%%%%%%%%%%%%%%%%%%%%%%%%%%%%%%%%%%%%%%%%%%%%%%%%%%%%%%%%%%%%
%%%%%%%%%%%%%%%%%%%%%%%%%%%%%%%%%%%%%%%%%%%%%%%%%%%%%%%%%%%%%%%%%%%%%%
%%%%%%%%%%%%%%%%%%%%%%%%%%%%%%%%%%%%%%%%%%%%%%%%%%%%%%%%%%%%%%%%%%%%%%
%%%%%%%%%%%%%%%%%%%%%%%%%%%%%%%%%%%%%%%%%%%%%%%%%%%%%%%%%%%%%%%%%%%%%%
\section{Conclusions}
In this paper we have shown rigorously the existence and uniqueness of the solution of the Hamiltonian constraint of the lowest-order effective action of the complete nonprojectable Ho\v{r}ava theory. The constraint is solved for the lapse function $N$ and it is subject to Dirichlet boundary conditions at spatial infinity. For the proof we have assumed a kind of positivity condition on the spatial metric and its conjugate momentum. We have shown that the Hamiltonian constraint is a PDE for the absolute value of the lapse function and, under the same positivity condition, we have obtained a condition of consistency: the equation preserves the non negativity of the absolute value. We have also found that the lapse function is bounded from above, reaching its maximum value at spatial infinity. Of course, it would be interesting to study the existence in the general case, without imposing the positivity condition on the spatial metric and its conjugate momentum.

The fact that in the complete Ho\v{r}ava theory, unlike general relativity, the lapse function is determined by other field variables allowed us to determine its asymptotic behavior at spatial infinity, finding the behavior $N = 1 + \mathcal{O}(r^{-1})$ in any of the allowed frames of the theory. As a consequence, the evolution equations yield $\dot{g} = \mathcal{O}(r^{-2})$, a (unexpected) behavior different to general relativity. In particular, this implies that the evolution equations preserve the mode of order $r^{-1}$ of the initial data $g_{ij}(t=0)$. In addition, the variational calculus in the complete Ho\v{r}ava theory must be defined such that $\dot{g} = \mathcal{O}(r^{-2})$ is preserved.

A nice feature of the theory we have found is that, after a conformal transformation, the Hamiltonian constraint can be solved in terms of the variable that exactly gives the energy of the theory. This is similar to the perturbative ADM Hamiltonian analysis of general relativity, where the Hamiltonian constraint is solved for the scalar mode of the transverse sector and the energy is defined by the asymptotic behavior of this mode. Using this conformal transformation we proved the positivity of the energy of the large-distance Ho\v{r}ava theory, under the assumption $G\geq 0$. The minimal energy is obtained when $N=1$, in this case the theory describes Minkowski spacetime. Although the Ho\v{r}ava theory is not invariant under Lorentz tansformations, Minkowski spacetime is recovered under the minimal energy condition.

%%%%%%%%%%%%%%%%%%%%%%%%%%%%%%%%%%%%%%%%%%%%%%%%%%%%%%%%%%%%%%%%%%%%%%
%%%%%%%%%%%%%%%%%%%%%%%%%%%%%%%%%%%%%%%%%%%%%%%%%%%%%%%%%%%%%%%%%%%%%%
%%%%%%%%%%%%%%%%%%%%%%%%%%%%%%%%%%%%%%%%%%%%%%%%%%%%%%%%%%%%%%%%%%%%%%
%%%%%%%%%%%%%%%%%%%%%%%%%%%%%%%%%%%%%%%%%%%%%%%%%%%%%%%%%%%%%%%%%%%%%%
\section*{Acknowledgement}
A.~R.~and A.~S.~are partially supported by Project Fondecyt 1121103, Chile.

%%%%%%%%%%%%%%%%%%%%%%%%%%%%%%%%%%%%%%%%%%%%%%%%%%%%%%%%%%%%%%%%%%%%%%
%%%%%%%%%%%%%%%%%%%%%%%%%%%%%%%%%%%%%%%%%%%%%%%%%%%%%%%%%%%%%%%%%%%%%%
%%%%%%%%%%%%%%%%%%%%%%%%%%%%%%%%%%%%%%%%%%%%%%%%%%%%%%%%%%%%%%%%%%%%%%
%%%%%%%%%%%%%%%%%%%%%%%%%%%%%%%%%%%%%%%%%%%%%%%%%%%%%%%%%%%%%%%%%%%%%%

\end{document}